\documentclass{llncs}
\usepackage{amssymb}
\usepackage{graphicx}
\newtheorem{fact}{Fact}
\usepackage[lined,boxed,commentsnumbered,ruled,vlined]{algorithm2e}
\newtheorem{observation}{Observation}
\title{Inplace Algorithm for Priority Search Tree and its use in
Computing Largest Empty Axis-Parallel Rectangle}
\input{psfig.sty}
\author{Minati De \and Subhas C. Nandy}
\institute{Indian Statistical Institute,  Kolkata - 700108, India.}
\begin{document}
\maketitle
\begin{abstract}
There is a high demand of space-efficient algorithms in built-in or
embedded softwares. In this paper, we consider the problem of
designing space-efficient algorithms for computing the maximum area 
empty rectangle (MER) among a set of points inside a rectangular 
region $\cal R$ in 2D. We first propose an inplace algorithm for
computing the priority search tree with a set of $n$ points in $\cal
R$ using $O(\log n)$ extra bit space in $O(n\log n)$ time. It supports
all the standard queries on priority search tree in $O(\log^2n)$ time.
We also show an application of this algorithm in computing the
largest empty axis-parallel rectangle. Our proposed algorithm
needs $O(n\log^2n +m)$ time and $O(\log n)$ work-space apart from the
array used for storing $n$ input points. Here $m$ is the number of
maximal empty rectangles present in $\cal R$. Finally, we consider the
problem of locating the maximum area empty rectangle of arbitrary
orientation among a set of $n$ points, and propose an $O(n^3\log n)$
time in-place algorithm for that problem. 
\end{abstract}

\section{Introduction}
\label{intro}
Though memory is getting cheap day by day, still there are massive
demand for the low memory algorithms for practical problems which need
to be run on tiny devices, for example, sensors, GPS systems, mobile
hand-sets, small robots, etc. Also, now-a-days the data available in
several problems itself is huge. So, the practical algorithms for the
data-streaming or data-mining problems must be space-efficient. For
all these reasons, designing low-memory algorithms for practical
problems have now become a challenging task in the algorithm research.

In computational geometry, designing the in-place algorithms are
studied only for a very few problems. For convex hull problem in both
2D and 3D, space efficient algorithms are available. In 2D, the best
known result is a $O(n\log h)$ algorithm with $O(1)$ extra space
\cite{B04}. Bronnimann et al. \cite{BCC04} also showed that the upper
hull of a set of $n$ points in 3D can be computed in $O(n\log^3n)$
time using $O(1)$ extra space. In the same paper it is also shown that 
for a parameter $s$ satisfying $c\log^2n \leq s \leq n$, ($c \geq 0$ 
is a constant), if $O(\frac{n}{s})$ space is allowed, then the convex 
hull for a set of $n$ points in 3D can be computed in in $O(ns)$ time.
Vahrenhold \cite{V07} proposed an $O(n^{\frac{3}{2}}\log n)$ time and
$O(1)$ extra space algorithm for the Klee's measure problem, where
the objective is to compute the union of $n$ axis-parallel rectangles
of arbitrary sizes. Bose et al. \cite{BMM07} used in-place divide and
conquer technique to solve the following three problems in 2D using
$O(1)$ extra space: (i) a deterministic algorithm for the closest pair
problem in $O(n\log n)$ time, (ii) a randomized algorithm
for the bichromatic closest pair problem in $O(n\log n)$ expected 
time, and (iii) a deterministic $O(n\log n + k)$ time algorithm for
the orthogonal line segment intersection computation problem. For
arbitrary line segments intersection computation problem, two
algorithms are available in \cite{CC03}. If the input array can be
used for storing intermediate results, then the problem can be solved
in $O((n+k)\log n)$ time and $O(1)$ space. but, if the input array is
not allowed to be destroyed, then the time complexity increases by a
factor of $\log n$, and it also requires $O(\log^2 n)$ extra space.
Regarding the empty space recognition problem, it needs to be
mentioned that all the Delauney triangles among a set of $n$ points
can be computed in $O(n^2)$ time using $O(1)$ space \cite{AR}.
This, in turn, recognizes the largest empty circle among a point 
set with the same time complexity.  

In this paper, we propose an in-place algorithm for constructing a
priority search tree $\cal T$ \cite{pst} with the set of points $P$
in $\cal R$, $|P|=n$. This needs an additional $O(\log n)$ bits. We
show that, the standard queries on priority search tree can be
performed in it in 
$O(\log^2n)$ time. Priority search tree is a very important paradigm
in geometric algorithms, as it has several important applications. 
Thus, our algorithm may aid in several problems in memory-restricted
environment. 

As an immediate application of our inplace priority search tree, we
have considered the computation of largest empty axis-parallel
rectangles among the points in $P$. Our proposed algorithm runs in
$O(m+n\log^2 n)$ time using $O(\log n)$ extra space. Here $m$ is the
number of maximal empty axis-parallel rectangles (MERs) in $\cal R$.
By maximal empty axis-parallel rectangle (MER), we mean an empty
axis-parallel rectangle that is not containined in any other empty
axis-parallel rectangle. 

The problem of computing the largest MER was first introduced by
Namaad et al. \cite{NHL}. They showed that the number of MERs' ($m$)
among a set of $n$ points may be $O(n^2)$ in the worst case; but if
the points are randomly placed, then the expected value of $m$ is
$O(n\log n)$. In the same paper, an algorithm for identifying the 
largest MER was proposed. The worst case time complexity of that 
algorithm is $O(min(n^2,m\log n))$. Orlowski \cite{OR} proposed an 
easy to implement algorithm for finding the largest MER that runs in 
$O(m + n\log n)$ time. It inspects all the MERs' present on the
floor, 
and identifies the largest one. The best known algorithm for this
problem runs in $O(n\log^2n)$ time in the worst case \cite{AS}. 
Same time 
complexity result holds for the recognition of the largest MER among 
a set of arbitrary polygonal obstacles \cite{NSB}. Recently, Boland 
and Urrutia \cite{BU} gave an $O(n\log n)$ time algorithm for finding 
the largest MER inside an $n$-sided simple polygon. All these 
algorithms use $O(n)$ extra space. 

Finally, we considered the problem of designing an 
in-place algorithm for computing the largest empty rectangle of 
arbitrary orientation among a set of $n$ points. It takes 
$O(n^3\log n)$ time with an $O(1)$ additional workspace. The best
known algorithm for this problem in the literature runs in $O(n^3)$
time and $O(n^2)$ space \cite{CND}.

\section{In-place priority search tree}
\label{bpst}
Let $P=\{p_1,p_2, \ldots, p_n\}$ be the given set of points in 2D, 
where $2^{k-1} \leq n < 2^k$. The array that stores the points 
in $P$, is also referred to as $P$. The $i$-th array location will be 
referred as $P[i]$. We now define the priority search tree $\cal T$ 
recursively in a way that suits our in-place implementation. 

The tree $\cal T$ has exactly $k$ levels. The level of root in $\cal T$ is 
considered to be the $k$-th level. The root represents the entire 
set of points $P$, and it stores the point $p^*$, where $y(p^*) = 
\max_{p\in P} y(p)$. Its two children are the priority search tree 
with the set of points $P_\ell$ and $P_r$, where $P_\ell$ and $P_r$ 
are defined as follows: (i) let $m= 2^{k-1}$. Compute $m$-th order 
statistics $x_{(m)}$ among the $x$-coordinates of the points of $P$. 
The set $P_\ell = \{p \in P \mid p \neq p^*, x_p \leq x_{(m)}\}$, 
and $P_r=\{p \in P \mid p \neq p^*, x_p > x_{(m)}\}$. Thus, 
$|P_\ell|=2^{k-1}-1$, $|P_r| =n-2^{k-1}$, and $P_\ell \cup P_r
= P\setminus\{p^*\}$. In our modified definition of priority search 
tree, we assume that each node at level $i$ $(i \neq 0$) represents 
a tree of size $2^i-1$ except the rightmost node in that level. 
Observe that, at the level $k-1$, each of the two subtrees of 
$P_\ell$ are of size $2^{k-2}-1$. But for $P_r$, we may not always 
be able to construct two subtrees if $|P_r| \leq 2^{k-2}$. In that 
case, it has only the left subtree rooted at the point having maximum 
$y$-coordinate in the point set $P_r$; otherwise it has two subtrees. 

In general, at any node of the $i$-th level, the root is as defined 
earlier. If $P'$ denotes the set of points represented by that node, 
then the number of children of that node is (i) zero or (ii) one or 
(iii) two depending on whether (i) $|P'|=1$, or (ii) $1 < |P'| \leq 
2^{i-1}$ or (iii) $|P'| > 2^{i-1}$. In Case (i), it has no subtree. 
In Case (ii), it has only the left subtree, and its size is $|P'|-1$. 
In Case (iii), it has both left and right subtrees, and their sizes 
are $2^{i-2}-1$ and $|P'|-2^{i-2}$ respectively. We maintain an 
array $TAG$ with $2\log n$ cells, indexed by the levels of the tree 
$\cal T$. Each cell is of size 2 bits. The $TAG[i]$ is set to 0 or 1 
or 2 depending on whether $\Theta_i$ = the number of nodes at level 
$i$ of $\cal T$ is equal to $2\Theta_{i-1}$ or $2\Theta_{i-1}-1$ 
or $2\Theta_{i-1}-2$. Note that, unlike the usual priority search tree 
\cite{pst}, at each node the discriminating $x$-value among the points 
in two subtrees is not stored. Here, the method of deciding the search 
direction from a node will be decided by observing the inorder 
predecessor and successor of that node in $\cal T$.

\subsection{In-place organization of $\cal T$}
$\cal T$ is organized in a heap like structure. All its leaves appear 
in the same level; but unlike heap, a non-leaf node of $\cal T$ may 
have zero or one or two child(ren). In a particular level $i$, at most one 
node may have less than two children, and if such a node exists, it is the 
right-most node in that level. The root of $\cal T$ is stored in 
$P[1]$; it has always two children, stored in $P[2]$ 
and $P[3]$ respectively. In general, if all the nodes in $\cal T$ 
have two children except the leaves, then the children of $P[j]$ 
reside at $P[2j]$ and $P[2j+1]$. But, since at most one node in a 
level of $\cal T$ is permitted to have less than two children, we 
use $TAG$ bits to compute the address of the children of 
a node. If a node at level $i$, and stored at $P[j]$, has two children, 
they are available at $P[2j-\sum_{\alpha={i+1}}^k TAG[\alpha]]$ and 
$P[2j-\sum_{\alpha={i+1}}^k TAG[\alpha]+1]$ respectively. If $P[j]$ 
corresponds to the right-most node at level $i$, and $TAG[i]=1$, then 
its only child resides at $P[2j-\sum_{\alpha={i+1}}^k TAG[\alpha]]$.

\subsection{Creation of $\cal T$}
We first initialize $TAG[i]=0$ for all $i=1,2,\ldots,k$, where $k = 
\lfloor \log n \rfloor$. The computation of $\cal T$ is done in a 
breadth-first manner. In other words, all the nodes in a particular 
level $i$ are computed prior to computing the nodes of level $i+1$, 
for all $i=1,2,\ldots,k$. We compute $p^* = P[j]$ (say), where 
$y(P[j])=\max_{i=1}^n y(P[i])$ as the root of $\cal T$, and store 
it in $P[1]$ by swapping $P[1]$ and $P[j]$. Next we sort 
$P\setminus\{p^*\}$ in an in-place manner with $O(n)$ data movement 
\cite{FG}. Its first $2^{k-1}-1$ elements form the set $P_\ell$ 
for the left subtree ${\cal T}_\ell$, and the remaining 
$n-2^{k-1}-1$ points form the set $P_r$ for the right 
subtree ${\cal T}_r$. Next, we identify $p_\ell^*$ and $p_r^*$, the 
roots of ${\cal T}_\ell$ and ${\cal T}_r$ (as defined for $\cal T$). 
The point $p_\ell^*$ (resp. $p_r^*$) is swapped with $P[2]$ (resp. 
$P[3]$). Again we sort the points in $P \setminus \{p^*, p_\ell^*, 
p_r^*\}$ to compute the nodes in the next level. Note that, at this 
level, both the children of $p_\ell^*$ exists; but the number of 
children of $p_r^*$ may be zero or one or two. If the number of children of 
$p_r^*$ is one or zero, $TAG[2]$ is set to 1 or 2. The same process continues 
up to the $k$-th level. Since, at each level, a sorting is involved, 
we have the following result:

\begin{lemma}
For a given set $P$ of $n$ points, the tree $\cal T$ can be 
constructed in  $O(n\log^2n)$ time.
\end{lemma}

\subsection{Traversal of $\cal T$} \label{traversal}
The traversal in $\cal T$ starts from its root (at $P[1]$). At each 
step, it moves towards one of its children. If $\cal T$ is full, 
the children of a node (point) stored at $P[j]$ (at level $i$ of 
$\cal T$) are available at $P[2j]$ and $P[2j+1]$. But, since 
$\cal T$ may not be full, we need to use $TAG$ bits attached to 
each level of $\cal T$ for the traversal. We maintain an integer 
location $\Delta$ during the traversal of $\cal T$. While moving from 
level $i$ to level $i+1$, we add $TAG[i]$ with $\Delta$. It is already 
mentioned in the earlier subsection that, the left (resp. right) child 
of the node $P[j]$ are available at location $2j-\Delta$ (resp. $2j+1-
\Delta$). Again, if $P[j]$ is the right-most node of a level of 
$\cal T$, it may have zero, one or two children, and this information 
is available at $TAG[i]$. If $P[j]$ is the right-most in its level $i$, 
and it has only one child, then the algorithm has to move towards its 
left child irrespective of the requirement (of moving towards left or 
right) in the algorithm.

\section{Standard queries on priority search tree}
We now show that the standard queries on the priority 
search tree \cite{pst} can be performed in $\cal T$ also 
in $O(\log^2n)$ time with $O(1)$ additional space.

\subsection{MinXInRectangle($x_0,x_1,y_1$)}
Three real numbers $x_0$, $x_1$ and $y_1$ are given. The objective 
is to find the point $p^*=(x^*,y^*) \in P$ with minimum $x$-coordinate 
among those points $p=(x,y)\in P$ satisfying $x_0 \leq x \leq x_1$ and 
$y \geq y_1$.

Since the $x$-coordinate of the partitioning line at each node of 
$\cal T$ is not stored as in \cite{pst}, the search direction from a 
node $p\in {\cal T}$ is decided by its inorder predecessor $p^-$ and 
inorder successor $p^+$. 

While executing this query with the interval $[x_0,x_1]$, we first 
find the {\it discriminant} node $\pi \in \cal T$ such that $x(\pi^-) 
< x_1$ and $x(\pi^+) > x_0$.  If $y(\pi) < y_1$, then report that the search can 
not output a feasible point satisfying the query; otherwise the search 
proceeds to answer the query. We initialize two locations $p^*$ and 
$\Delta^*$ with $\pi$ and $\Delta$, where $p^*$ will contain the final 
answer, and $\Delta$ is the sum of $TAG$ bits computed during the 
traversal up to the node $\pi$. The search proceeds in both the subtrees of 
$\pi$. The traversal in the left subtree of $\pi$ starts with $p=\pi$. 
The actions taken at each node $p$ on the search path, and the choice 
of its child for the next move is decided as follows.

\begin{itemize}
\item[$\bullet$] If $y(p) < y_1$, the search stops. Otherwise,  
execute the following three steps.
\item[$\bullet$] If $x_0 \leq x(p) \leq x_1$ then assign 
$p^* = p$
\item[$\bullet$] If the $x(p^-) < x_0$, then set $p$ = right child of $p$.
\item[$\bullet$] If the $x(p^+) > x_0$, then set $p$ = left child of $p$ 
\end{itemize}

Using a similar procedure we traverse the right subtree of $\pi$ 
starting with $p=\pi$, and restoring the value of $\Delta$ at node 
$\pi$, which is stored in $\Delta^*$. Finally, $p^*$ is reported 
as the answer to the query. 

{\bf Time complexity:} Since here the search direction from a node 
$q$ is guided by $x(q^-)$ and $x(q^+)$, each move from a node to its descendant in the direction of the search takes 
$O(\log n)$ time. Note that, while, searching $q^-$ or $q^+$ of a 
node $q$, we copy $\Delta$ of $q$ at a scalar location $\Delta'$, 
and perform the search as mentioned in subsection \ref{traversal}. 

Since the total number of nodes to be traversed to report the answer 
or the non-existence of a feasible solution is $O(\log n)$, we have 
the following lemma:

\begin{lemma}
Using the in-place priority search tree on a set of $n$ points, the 
{\bf MinXInRectangle}($x_0,x_1,y_1$) query can be answered in 
$O(\log^2n)$ time using $O(1)$ extra space. 
\end{lemma}

\subsection{MaxXInRectangle($x_0,x_1,y_1$)}
Three real numbers $x_0$, $x_1$ and $y_1$ are given. The objective is 
to find the point $p^*=(x^*,y^*) \in P$ with maximum $x$-coordinate 
among those points $p=(x,y) \in P$ satisfying $x_0 \leq x \leq x_1$ 
and $y \geq y_1$. This query can be answered in a similar manner as 
in {\bf MinXInRectangle} query with same time and space complexity.

\subsection{MaxYInXRange($x_0,x_1$)}
Given a pair of real numbers $x_0$ and $x_1$, find a point $p^*=
(x^*,y^*)$ whose $y$ coordinate is maximum among all the points in 
$P$ satisfying $x_0\leq x\leq x_1$. Here the algorithm is essentially 
the same as in {\bf MinXInRectangle} query. Here if a node satisfies 
$x_0 \leq x \leq x_1$ during the search for the discriminant node, 
the search stops reporting that point. Otherwise, we need to search 
separately both the subtrees of the discriminant node till a point is 
obtained satisfying $x_0 \leq x \leq x_1$. Surely, time and space 
complexities of {\bf MinXInRectangle} query hold here also.

\subsection{EnumerateRectangle($x_0,x_1,y_1$)}
Three real numbers $x_0$, $x_1$ and $y_1$ are given. The objective 
is to identify all the points $p=(x,y) \in P$ satisfying $x_0 \leq 
x \leq x_1$ and $y \geq y_1$. Here, the {\it discriminant} point $\pi$ 
is found as in {\bf MinXInRectangle} query. In this path, if there 
exists any point satisfying the given constraint, then report it. 
Next, perform an inorder traversal in the subtree rooted at $\pi$ 
to identify all the points satisfying the desired condition. During 
the inorder traversal, (i) if a node with inorder predecessor having 
$x$-coordinate less than $x_0$ is reached, its left subtree is not 
traversed, similarly (ii) if a node with inorder successor having 
$x$-coordinate greater than $x_1$ is reached, its right subtree is 
not traversed, and (iii) if a node is reached whose $y$-coordinate 
is less than $y_1$, then its both the subtrees are not traversed. 

Also note that, from a node $P[j]$ at level $i$, the index of 
its parent (at level $i+1$) in the array $P$ is computed as 
$\lfloor\frac{j+TAG[i-1]}{2}\rfloor-TAG[i-1]$. At each movement from 
a node at level $i$ to its parent at level $i+1$ in $\cal T$, we need to 
update $\Delta$ by $\Delta-TAG[i]$. Thus we have the complexity 
results of this query as stated below:
\begin{lemma}
Using the in-place priority search tree on a set of $n$ points, the 
{\bf EnumerateRectangle}($x_0,x_1,y_1$) query can be answered in 
$O(m+\log^2n)$ time using $O(1)$ extra space, where $m$ is the size 
of the reported answer. 
\end{lemma}

\section{Largest empty rectangle}
\label{ler}
We now concentrate on our main problem of computing the largest 
empty axis-parallel rectangle among the point set $P$ stored in an 
axis-parallel rectangular region $\cal R$. The 
algorithm consists of two phases: top-down and bottom-up. Each 
phase consists of $n$ passes. In each pass of the top-down phase, 
we identify all the MERs with the point stored at the root node 
of $\cal T$, on its top boundary, and then delete the root from 
$\cal T$. Thus, a new point having maximum $y$-coordinate among 
the remaining points in $P$ becomes the root. The same algorithm 
is repeated to compute MERs with the new root at their top boundary. 
The deletion of the root of $\cal T$ is explained in detail. After 
the deletion, a space in $\cal T$ becomes empty. Actually, we store 
the deleted root of $\cal T$ in that location. We show that, it 
does not affect the correctness of our algorithm. Thus, after the 
completion of the top-down phase, all the points in $P$ are present 
in the array $P$, and we can execute the bottom-up phase with all 
the points in $P$ stored in the same array. The bottom-up phase is 
exactly similar to the top-down phase. Here, in each pass, all the 
MERs with bottom boundary passing through the point stored in the 
root node of $\cal T$ are identified. We now explain the top-down 
phase in detail.

\subsection{Top-down phase}
We now explain the processing of the root $p^*=P[1]$ of $\cal T$. 
Let $x_{min}$ and $x_{max}$ be the left and right boundary of 
$\cal R$ respectively. We use two double-ended queues $Q_\ell$ and 
$Q_r$ to store the points encountered in the two sides of $p^*$ 
during the traversal of $\cal T$. It stores some already visited 
points of $\cal T$ for the future processing, and will be clear 
from subsequent discussions. The insertion and deletion in both 
the queues can be performed in both of their ends. At the begining 
of each pass, $Q_\ell$ and $Q_r$ are empty. We define a curtain with 
horizontal span ${\cal I} = [x_{min}, x_{max}]$; its top boundary is 
fixed at $p^*$. Let $p_\ell$ and $p_r$ be the two children of $p^*$. 
If $p_\ell$ and $p_r$ are in 
different sides of $p^*$, then both the points are pushed in their 
respective queues. But, if $p_\ell$ and $p_r$ are in the same sides 
(say left) of $p^*$, then two situations need to be considered: (i) 
if $y(p_\ell) > y(p_r)$, then both $p_r$ and $p_\ell$ are pushed in 
$Q_\ell$ in this order. Otherwise, $p_r$ is pushed in $Q_\ell$, and 
$p_\ell$ is ignored.

Next time onwards, the point $p$ for the processing is the one having 
higher $y$-coordinate among the front elements of $Q_\ell$ and $Q_r$.
While choosing the point $p$ for processing, it is first deleted from 
the respective queue, and then the MER is reported as stated below. 
It also causes updating of the queues $Q_\ell$ and $Q_r$.

\subsection{Processing the topmost queue element}
Let ${\cal I}=[\alpha,\beta]$ denote the horizontal span of the 
curtain, We first report an MER with the horizontal span $\cal I$, 
and vertical span $[y(p^*),y(p)]$. $\cal I$ is truncated at $p$, 
so that $p^*$ lies in the updated $\cal I$. Here also, we 
will use $p_\ell$ and $p_r$ to denote the children of $p$. Depending 
on the position of $p_\ell$ and $p_r$ with respect to $p^*$ and $p$, 
one or both of $p_\ell$ and $p_r$ are put in the appropriate queue 
as described below. 

\begin{figure}[t]
\centering
\includegraphics[scale=0.4]{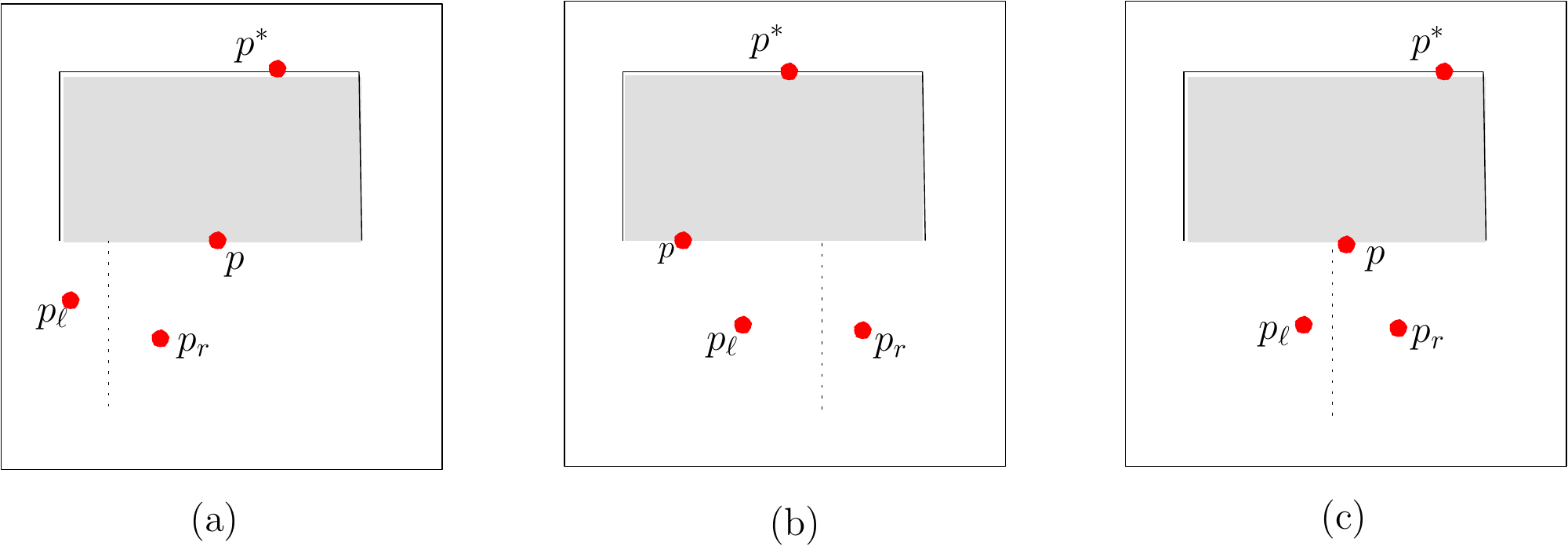}
\caption{Three different cases while processing a point $p$ in 
the top-down phase} 
\label{fig1}
\end{figure}

Without loss of generality, let us assume that $p$ is to the left of
$p^*$, and $p_\ell$ and $p_r$ be its two children. Here, three cases
may arise: (i) both $p_\ell$ and $p_r$ are to the left of $p$ 
(Figure \ref{fig1}(a)), (ii) both $p_\ell$ and $p_r$ are to the right of 
$p$ (Figure \ref{fig1}(b)), and (iii) $p_\ell$ and $p_r$ are in 
different sides of $p$ (Figure \ref{fig1}(c)). The actions taken in 
the three different cases are stated below.

\begin{figure}[h]\vspace{-0.1in}
\centering
\includegraphics[scale=0.4]{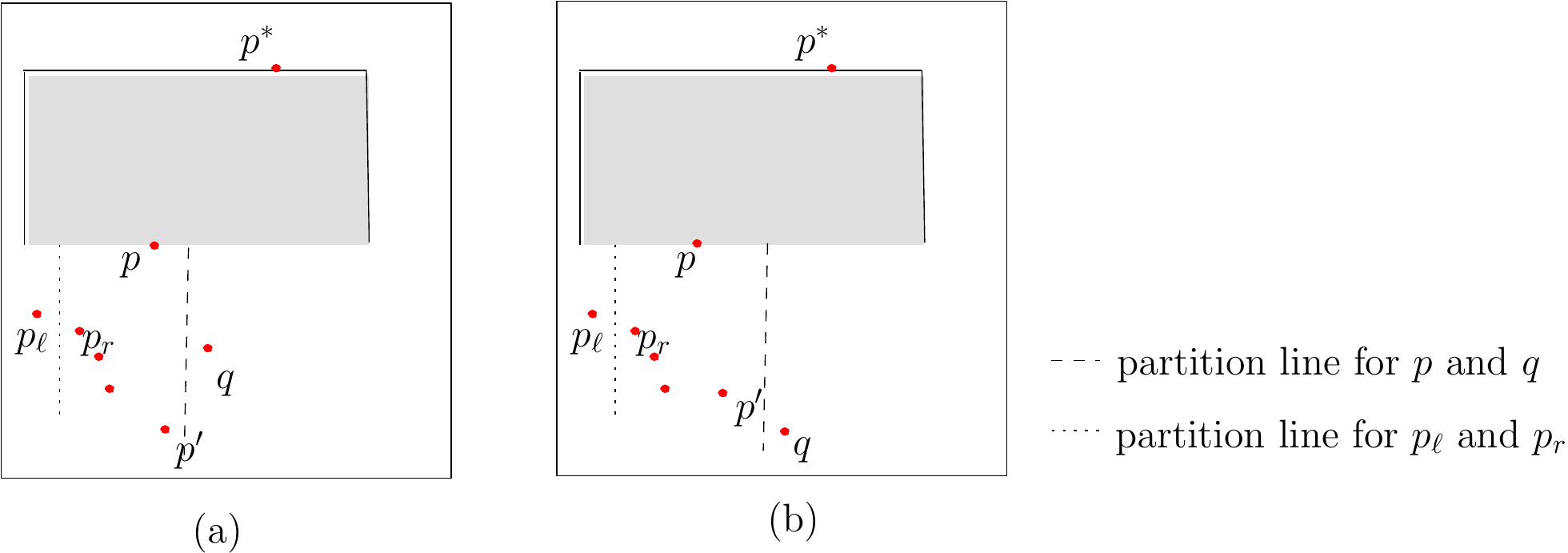}\vspace{-0.1in}
\caption{Demonstration of Case (i)} \vspace{-0.1in}
\label{fig2}
\end{figure}

\begin{description}
\item[Case (i)] Here, both $p_\ell$ and $p_r$ are not inserted in $Q_\ell$. 
However, we need to follow the right links starting from
$p_r$ until (a) a point $p'$ is found inside $\cal I$, or (b) the 
bottom boundary of the floor is reached (i.e. the index of the right 
child of the current node on the traversal path is greater than the size $n$
of the array $P$). 
In case (a), surely we have $x(p') < x(q)$ where $q$ is the element 
at the front-end of $Q_\ell$\footnote{The reason is that the point 
is $p'$ is in the same partition of $p$ and the point $q$ is entered 
in $Q_\ell$ by the right sibling of $p$ or the right child of some 
ancestor of $p$.}. Now, if $y(p')> y(q)$, we insert $p'$ at the 
front-end of $Q_\ell$ (see Figure \ref{fig2}(a)). But, if $y(p') < 
y(q)$, we ignore $p'$, or in other words, do not insert it in $Q_\ell$ 
(see Figure \ref{fig2}(b)). 

\begin{figure}[h]
\centering
\includegraphics[scale=0.4]{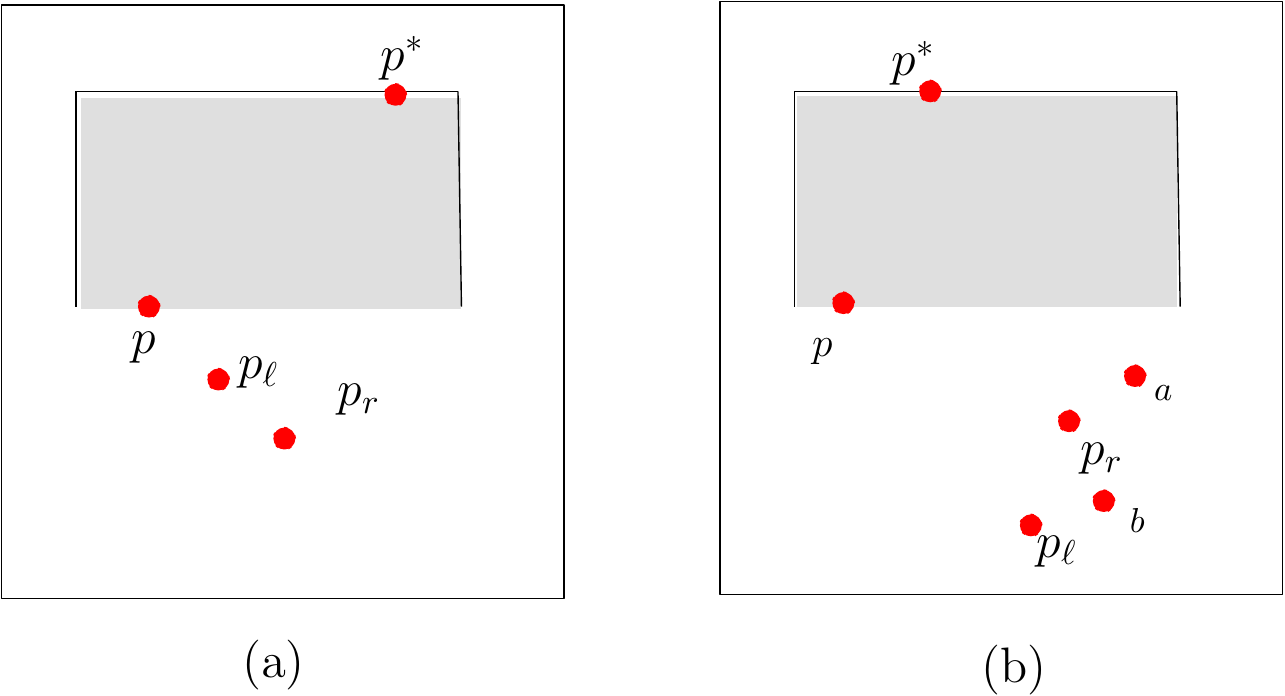}
\hspace{0.2in}
\includegraphics[scale=0.4]{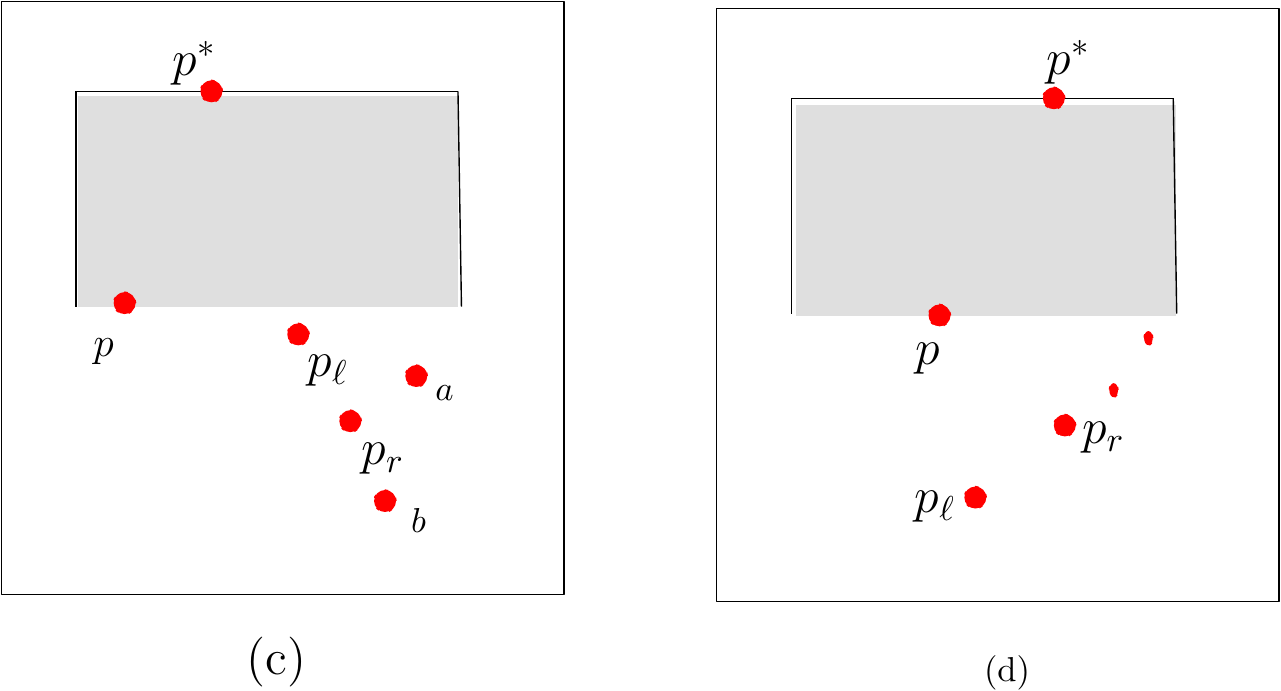}
\vspace{-0.1in}
\caption{Demonstration of Case (ii)} 
\label{fig3}
\end{figure}

\item[Case (ii)] Here three different situations may arise: (a) both 
$p_\ell$ and $p_r$ are to the left of $p^*$, (b) both $p_\ell$ and 
$p_r$ are to the right of $p^*$, and (c) $p_\ell$ and $p_r$ are in 
different sides of $p^*$. In subcase (a), if $y(p_\ell) > y(p_r)$, we 
insert $p_r$ and $p_\ell$ at the front-end of $Q_\ell$ in this order 
(see Figure \ref{fig3}(a)); otherwise only $p_r$ is put in $Q_\ell$, 
and $p_\ell$ is ignored. In subcase (b), we insert both $p_r$ and 
$p_\ell$ at the rear-end of $Q_r$. This may need deleting elements of 
$Q_r$ from its rear-end. Such a situation is demonstrated in Figure 
\ref{fig3}(b). Here $a$ and $b$ are already present in $Q_r$; while inserting $p_r$, 
$b$ is deleted from its rear-end since $y(b) < y(p_r)$. Next, $p_\ell$ 
is also inserted in $Q_r$ in the same manner. Note that, while 
inserting $p_\ell$, $p_r$ may also be deleted (see Figure 
\ref{fig3}(c)). In subcase (c), $p_\ell$ and $p_r$ will need to be 
inserted in $Q_\ell$ and $Q_r$ respectively. Note that, if $Q_\ell$ 
is non-empty, then $p_\ell$ appears to the left of all the elements
present in $Q_\ell$. Thus, if $y(p_\ell)$ is greater than the 
$y$-coordinate of the element present at the front-end of $Q_\ell$, 
then $p_\ell$ is inserted at the front-end of $Q_\ell$. Similarly, 
if there is any element in $Q_r$, $p_r$ will be left to all of them. 
Here, $p_r$ must be added at the rear-end of $Q_r$; if needed, some 
element of $Q_r$ are deleted from its rear-end. The tiny points 
in Figure \ref{fig3}(d) are already present in the $Q_r$. Note that, 
if this situation arises, at least one of $Q_\ell$ and $Q_r$ 
must be empty. 

\begin{figure}[h]
\centering
\includegraphics[scale=0.4]{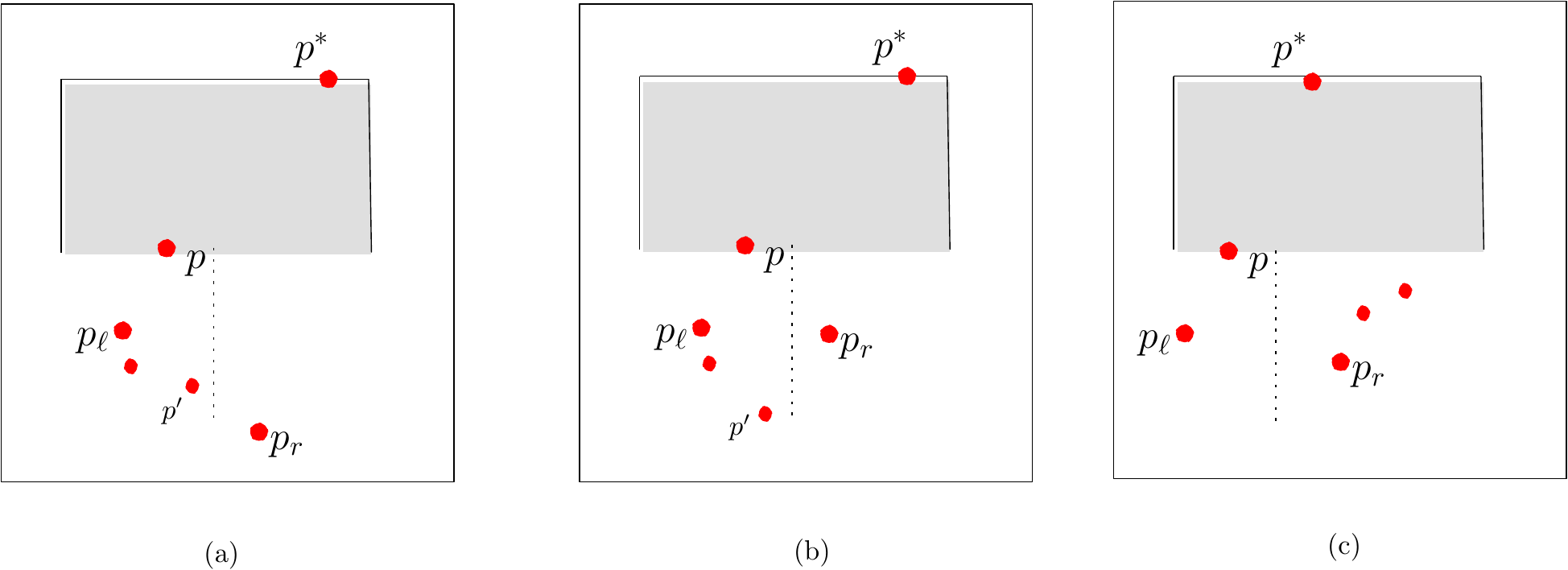}\vspace{-0.15in}
\caption{Demonstration of Case (iii)} 
\label{fig4}
\end{figure}

\item[Case (iii)] Here, first $p_r$ is inserted in either at the 
front-end of $Q_\ell$ or at the rear-end of $Q_r$ depending on whether
$p_r$ is to the left or right of $p^*$ (see Figure \ref{fig4}). Note 
that, here $p_\ell$ must be ignored. But as in case (i), we need to 
proceed following the right links starting from $p_\ell$ to reach 
either a point $p'$ inside $\cal I$ or the bottom boundary of the 
floor. If $y(p') > y(q)$, where $q$ is at the front-end of $Q_\ell$, 
then $p'$ is entered in $Q_\ell$ at its front-end (see Figure 
\ref{fig4}(a)); otherwise $p'$ is ignored (see Figure \ref{fig4}(b)). 
Figure \ref{fig4}(c), shows a situation where $p_r$ is to be inserted 
at the rear-end of $Q_r$. 
\end{description}

If $p$ has a single child, it is inserted in the queue in a similar 
manner. If $p$ has no child, no special action needs to be taken; we 
only proceed to process the next point in the queue. 

The similar set of actions are taken for processing the point $p$ when 
it is at the right side of $p^*$. The execution continues until both 
the queues become empty. Then the last (only one) MER is reported with 
the resulting curtain $\cal I$ as the horizontal span, and vertical span 
defined by $p^*$ and the bottom boundary of $\cal R$.

\subsection{Deletion of the root of $\cal T$}
After the completion of a pass, we reorganize the tree $\cal T$ by 
deleting the point at its root as follows, without computing it afresh. 

We start from the root by assigning $i=1$. At each 
move we consider both the children of $P[i]$, and choose the one, say 
$P[j]$ having maximum $y$-coordinate. The tie, if arises, is broken 
arbitrarily. If $P[i]$ has only one child, its index is taken in $j$. 
We swap $P[i]$ and $P[j]$. and the algorithm proceeds by copying the 
value of $j$ in $i$. Finally, when no child of $P[i]$ is found, the 
algorithm terminates. Here the following facts need to be mentioned:
\vspace{-0.1in}
\begin{fact}\label{fact1}
The point at the root is logically deleted, but it still remains in 
the array $P$. This is essential, since we need to execute a 
bottom-up phase with all the points in $P$ after the execution of 
the top-down phase.
\end{fact}
\begin{fact}\label{fact2}
Usually, while traversing along a path of $\cal T$, the bottom of 
the floor $\cal R$ is recognized, when a leaf is reached, or in other 
words, the index of the child of the said node along the desired 
direction is greater than the array size $n$. But, such a method may 
fail from the second pass onwards due to our scheme of deleting the 
root. Here apart from the usual way, the leaf is also detected if a 
node (point) is reached whose $y$-coordinate is greater than the 
$y$-coordinate of the root of $\cal T$. 
\end{fact}
\begin{fact}\label{fact3}
During the deletion of the root of $\cal T$, when 
a point is moved from a child node to its parent, it still remains in 
its own partition according to the partition line defined at that node 
at the time of creation of the $\cal T$. Thus, binary search property 
according to the $x$-coordinate values in the present $\cal T$ still 
remains valid. Moreover, the point having maximum $y$-coordinate in a 
partition is at the root of its corresponding subtree. Thus the 
modified $\cal T$ is still a priority search tree.
\end{fact}
\begin{fact}\label{fact4}
The tree $\cal T$ may no longer remain balanced after the deletion of 
some points from $\cal T$. Or in other words, leaf node may appear in 
different level. However, the length of each search path will still be 
bounded by $O(\log n)$.
\end{fact}
\vspace{-0.1in}
By the virtue of Facts \ref{fact1}, \ref{fact2}, \ref{fact3}, we can 
execute the subsequent passes to identify all the MERs with top boundary 
aligned at the point staying at the root node of $\cal T$ in that pass. 
Fact \ref{fact4} says that the time required for deletion of root at 
the end of each pass is $O(\log n)$.

%\vspace{-0.2in}
\subsection{Complexity Analysis}\vspace{-0.1in}
\begin{lemma}\label{lx}
A single pass of the top down phase needs $O(\mu+\log n)$ time in the 
worst case, where $\mu$ is the number of reported answers in that pass. 
\end{lemma}
\vspace{-0.1in}
\begin{proof}
In each pass, when an MER is reported, at most two points are inserted 
in the queue. Thus, the number of points inserted in the queue is at most 
$2\mu$. Though the insertion of a point in the front-end always 
takes $O(1)$ time, the insertion of a point in the rear-end may need some 
deletions prior to that. But, since the total number of deletions in a pass 
is at most equal to the number of insertions in that pass, and no point 
is inserted twice in a 
pass, the time required for a single pass in the top-down phase is 
$O(\mu)$. The time for the deletion of the root at the end of a pass 
needs $O(\log n)$ time. Thus, the result follows. \qed
\end{proof} 
\vspace{-0.1in}
\begin{lemma}\label{ly}
$|Q_\ell|$ and $|Q_r|$ can be at most $O(\log n)$ 
at any instant of time.
\end{lemma}
\vspace{-0.1in}
\begin{proof}
The result follows from the fact that at any instant of time, $Q_\ell$ 
(resp. $Q_r$) contains at most two points of a particular level in
$\cal T$. We justify this claim assuming the contradiction. Let $\Pi$ ($|\Pi| > 2$) be the set of 
points at the same level of $\cal T$ that are present in $Q_\ell$ at 
an instant of time. Surely, the parents for all of them are not the 
same, and they are inserted when their parents produced MERs. If we 
consider the $x$-coordinates of these points as well as their parents, 
there may exist at most two points (more specifically, the right-most 
two points) in $\Pi$, say $\pi_1$ and $\pi_2$, such that $x(\pi_1) > 
x(\pi)$ and $x(\pi_2) > x(\pi)$, where $\pi$ is the right-most one 
among the already processed parent nodes. Moreover, if such a 
situation arises, then $\pi_1$ and $\pi_2$ are the children of $\pi$. 
Thus, the other points of $\Pi$ have $x$-coordinate less than $x(\pi)$, 
and hence they can not belong in $Q_\ell$. \qed
\end{proof}
\vspace{-0.1in}
\begin{theorem}
The time complexity of our algorithm for identifying the maximum 
area/perimeter axis-parallel rectangle among a set of $n$ points is 
$O(m+n\log^2n)$, and it uses $O(\log n)$ work-space apart from the 
array $P$ containing input points.
\end{theorem}

\vspace{-0.1in}
{\bf Note:} The time complexity of the algorithm can be reduced to 
$O(m+n\log n)$ if $\cal T$ can be constructed for the first time in 
$O(n\log n)$ time avoiding the sorting at every level. 

\vspace{-0.2in}
\section{Finding MER of arbitrary orientation}\vspace{-0.15in}
We now propose an in-place algorithm for finding maximum area empty 
rectangle of arbitrary orientation among a set of points in  
$P$. The problem was addressed by Chaudhuri et al. \cite{CND}. They 
introduced the concept of PMER; it is the maximum area empty 
rectangle of any arbitrary orientation whose four sides are bounded 
by the members of $P$. It is shown that the number of PMERs is 
bounded by $O(n^3)$ in the worst case. It follows from the following 
observation:
\vspace{-0.1in}
\begin{observation} \label{obj1} \cite{CND}
At least one side of a PMER must contain two points from $P$, 
and other three sides either contain at least one point of $P$ 
or the boundary of $\cal R$. A corner incident at the boundary 
of $R$ implies that the corresponding sides contain that 
boundary point. . 
\end{observation}
\vspace{-0.1in}
The worst case time complexity of the algorithm proposed in 
\cite{CND} is $O(n^3)$, and it takes $O(n^2)$ work-space 
for executing the algorithm. Our algorithm uses $O(1)$ 
work-space but its worst case time complexity is $O(n^3\log n)$.

\vspace{-0.2in}
\subsection{Algorithm}\vspace{-0.1in}
Observation \ref{obj1} plays the central role in our algorithm. 
We consider each pair of points $p_i,p_j \in P$, and compute all 
the PMERs with one boundary passing through $p_i$ and $p_j$. We 
assume that the points in $P$ are increasingly ordered with 
respect to their $x$-coordinates; if tie occurs, then those points 
are increasingly ordered with respect to their $y$-coordinates. 
Two variables $i$ and $j$ are used to indicate the pair of points 
chosen for the processing. We choose different values of $(i,j)$, 
$i=1,2,\ldots,n-1$ and $j=i+1,i+2, \ldots,n$ in this order. Each 
time the pair $(i,j)$ is chosen, $p_i$ and $p_j$ are swapped with 
$p_1$ and $p_2$ respectively. 

We execute the procedure {\bf Process}$(p_i,p_j)$ to compute 
all the PMERs with their one boundary passing through $(p_i,p_j)$. 
Note that, after the execution of  {\bf Process}$(p_i,p_j)$, the 
points in $P$ will not be in the increasing order of their 
coordinates as mentioned above. Thus, in order to choose the next 
pair for the processing, we need to sort the array $P$ again with 
respect to their coordinates. 

\vspace{-0.1in}
\subsubsection{Process$(p_i,p_j)$:}\vspace{-0.1in}
Consider the straight line $\ell_{ij}$ passing through $p_i,p_j$. 
It is truncated by the boundary of $\cal R$ at its two ends. The 
points $p_i$ and $p_j$ are assumed to be stored in $P[1]$ and 
$P[2]$ respectively; the other points are split into two subsets 
according to the side of $\ell_{ij}$ it belongs. If $P_1$ and $P_2$ 
be the sets of points that are below and above $\overline{\ell_{ij}}$ 
respectively ($|P_1|+|P_2|=n-2$), then the points in $P_1$ are stored 
in $P[j], j=3,\ldots, |P_1|+2$, and the points of $P_2$ are stored in 
$P[j], j=|P_1|+3, \ldots, n$. We use the following procedure to 
partition $P\setminus\{p_i,p_j\}$ into $P_1$ and $P_2$. 

\vspace{-0.15in}
\begin{description}
\item[] Traverse the array from two directions using two index 
variables $\alpha$ and $\beta$, initialized to 3 and $n$. The 
variable $\alpha$ increases until a point in $P_2$ is observed; 
then $\beta$ starts decreasing until a point in $P_1$ is observed. 
Now, $P[\alpha]$ and $P[\beta]$ are swapped. The same process 
continues until $\alpha\leq \beta$. 
\end{description}\vspace{-0.15in}

We use two scalar locations $n_1$ and 
$n_2$ to store $|P_1|$ and $|P_2|$. We now sort both the set of points 
$P_1$ and $P_2$ separately with respect to their distances from 
$\ell_{ij}$. Note that, the distance values are not stored. While 
comparing a pair of points, their distance values are computed for 
the comparison.  We now describe the method of computing all the 
PMERs with the points in $P_1$, keeping $(p_i,p_j)$ at its top boundary. 

As in the earlier algorithm, we consider a curtain whose two sides are 
bounded by the boundary of $\cal R$, and top boundary is attached to 
both $p_i$ and $p_j$. The curtain is allowed to fall. As soon as it 
hits a point $p \in P_1$ it reports a PMER. This point can easily be 
obtained from the sorted list of $P_1$, stored in the array $P$. If 
the projection $\pi$ of the point $p$ on $\ell_{ij}$ lies inside the 
interval $[p_i,p_j]$, the processing of $\ell_{ij}$ stops. Otherwise, 
the curtain is truncated at $\pi$, and the 
process continues to process the next point in $P_1$. After 
finishing the processing of all the points in $P_1$, we process the 
points in $P_2$ in a similar manner to generate the PMERs with 
their bottom boundary passing through $p_i$ and $p_j$. 

\vspace{-0.15in}
\subsection{Correctness and complexity analysis}\vspace{-0.1in}
The correctness of the algorithm follows from the fact that 
for each pair of points $p_i,p_j\in P$, we have considered 
all possible PMERs with $(p_i,p_j)$ on its one side, and we 
have considered each pair of points in $P$.
  
Regarding the time complexity, note that, we have considered 
$O(n^2)$ pairs of points. For each pair, we have executed the 
procedure {\bf Process}. Each time after the execution of the 
procedure {\bf Process}, we need to sort the 
array $P$ with respect to their $x$ and $y$ coordinates for 
choosing the next pair of points for processing. 

In the procedure {\bf Process}, the splitting of $P$ into $P_1$ 
and $P_2$ needs $O(n)$ time. Sorting the members of 
$P_1$ and $P_2$ with respect to their distances from 
$\ell_{ij}$ needs $O(n\log n)$ time, and then the reporting of 
the PMERs need $O(n)$ time. 

Note that, we have only used four index variables $i$, $j$, 
$\alpha$ and $\beta$, and two integer locations $n_1$ and 
$n_2$ to store size of $P_1$ and $P_2$. Thus we have the 
following theorem stating the complexity results of our 
proposed algorithm. 
\vspace{-0.1in}

\begin{theorem}
Given a set of $n$ points, the maximum area/perimeter rectangle 
of arbitrary  orientation can be computed in $O(n^3\log n)$ 
time with $O(1)$ extra space. 
\end{theorem}

\end{document}